\documentclass[a4paper,11pt]{article}
\usepackage{pos}
\usepackage{slashed}
\usepackage{siunitx}

\title{ANUBIS: Projected Sensitivities and Initial Results from the proANUBIS demonstrator with Run 3 LHC data}

\author*[a]{Théo Reymermier}
\author[b]{Oleg Brandt}
\author[b]{Anna Mullin}
\author[b]{Paul Swallow}
\author[b]{Michael Revering}
\author[b]{Cayetano Fernandez Ruiz}
\onbehalf{for the ANUBIS Collaboration}

\affiliation[a]{Universit\'e de Lyon, Universit\'e Claude Bernard Lyon 1, CNRS/IN2P3, Institut de Physique des 2 Infinis de Lyon, UMR 5822,\\
F-69622, Villeurbanne, France}

\affiliation[b]{Department of Physics, Cavendish Laboratory, University of Cambridge,\\
JJ Thomson Avenue, Cambridge, CB3 0HE, United Kingdom}

\emailAdd{theo.reymermier@cern.ch}
\emailAdd{oleg.brandt@cern.ch}
\emailAdd{anna.jane.mullin@cern.ch}
\emailAdd{paul.nathaniel.swallow@cern.ch}

\abstract{Despite the success of the Standard Model (SM) there remains behaviour it cannot describe, in particular the presence of non-interacting Dark Matter. Many models that describe dark matter can generically introduce exotic Long-Lived Particles (LLPs). The proposed ANUBIS experiment is designed to search for these LLPs within the ATLAS detector cavern, located approximately 20-30 m from the Interaction Point (IP). A prototype detector, proANUBIS, has taken data within the ATLAS detector cavern since 2024, corresponding to 104 $fb^{-1}$ of pp data. We report on the potential sensitivity of ANUBIS to a selection of LLP models, i.e. Higgs Portal and Heavy Neutral Leptons, as well as future planned studies. Additionally, we will show the first results of the proANUBIS demonstrator, and how it will be used to study the expected backgrounds for the ANUBIS detector.}

\FullConference{The European Physical Society Conference on High Energy Physics (EPS-HEP2025)\\
7-11 July 2025\\
Marseille, France\\}

\begin{document}
\maketitle

\section{Introduction}

Long-Lived Particles (LLPs) are a key element for Beyond the Standard Model (BSM) physics, with connections to dark matter \cite{Hall:2010jx}, neutrino masses \cite{Bondarenko:2018ptm} and baryogenesis. Existing detectors, like ATLAS or CMS, lose acceptance for neutral LLPs with decay lengths above $\mathcal{O}(10\, m)$ \cite{ATLAS:2013dyx}. ANUBIS is a proposal for a transverse detector \cite{Bauer:2019vqk}, which would be situated in the ATLAS cavern, about 23 m above IP 1 and designed to observe traces of neutral LLPs that traverse ATLAS invisibly and decay into charged particles in direction of the cavern ceiling. Its large fiducial volume and timing synchronisation with ATLAS provide unique sensitivity and complementarity to proposals such as MATHUSLA \cite{Alpigiani:2018} and CODEX-b \cite{Aielli_2020}.

This paper presents projected sensitivities for two benchmark models, a Higgs-portal scalar and a minimal single-flavour Heavy Neutral Lepton (HNL). We further introduce the new \textsc{SET-ANUBIS} framework, which automates sensitivity studies for generic BSM models. Finally, we present the first result with the proANUBIS prototype, taking data in the cavern at ATLAS since 2024.

\section{LLP phenomenology and benchmark models}

\subsection{Generic LLP kinematics}

\label{sec:theory}
An LLP of width $\Gamma$ has $c\tau=\hbar c/\Gamma$ and lab decay length $L_\mathrm{lab}=\beta\gamma c\tau$. The probability to decay between $L_{\min}$ and $L_{\max}$ along its flight direction is
\begin{equation}
P_{\rm decay}=e^{-L_{\min}/(\beta\gamma c\tau)}-e^{-L_{\max}/(\beta\gamma c\tau)}.
\end{equation}
Signal expectations follow
\begin{equation}
N_{\rm sig}=\mathcal{L}\sum_i \sigma_i\,\mathrm{BR}_i\,
\langle A_i\,\varepsilon_i\,P_{\rm decay}\rangle,
\end{equation}
where $\langle A\varepsilon\rangle$ is the acceptance times the efficiency, $\mathcal{L}$ the luminosity and $\sigma_i$/$BR_i$ the production's cross section/branching ratio of the LLP for a given channel.

\subsection{Higgs portal} 

We consider a real singlet $S$ mixed with the SM Higgs $H$ through a mixing parameter $\epsilon$. This model is known as the Hidden Abelian Higgs Model (HAHM) \cite{Beacham_2019}. The Lagrangian is

\begin{equation}
    \mathcal{L}_{higgs} = -\frac{\epsilon}{2}S^2|H|^2 + \frac{\mu_S}{2}S^2 - \frac{\lambda_S}{4!}S^4 + \mu_H^2|H|^2-\lambda_H |H|^4 ,
\label{higgslagrangien}
\end{equation}

where $\mu_H/\lambda_H$ and $\mu_S/\lambda_S$ are the classical parameters for Higgs potentials. After ESWB, we get a singlet s of mass $m_s$ interacting with the Higgs (h). Production is via SM Higgs processes with $\mathrm{BR}(h\to ss)$; $s$ decays proceed through its Higgs admixture. For $m_s<2m_W$ (with $m_W$ the mass of the W boson),
\begin{equation}
\Gamma(s\to f\bar f)=\sin^2\!\theta\,\frac{N_c m_f^2 m_s}{8\pi}\left(1-\frac{4 m_f^2}{m_s^2}\right)^{3/2}.
\end{equation}
Here $N_c=3$ for quarks and 1 for leptons, $m_f$ the mass of the fermion and $\sin{\theta}$ is define in \cite{Curtin_2014}.Varying $(m_s,\sin\theta)$ spans proper lifetimes from sub-metre to cavern scales. We consider production through gluon-gluon fusion (54.7 pb). Vector-boson fusion (4.28 pb), Higgsstrahlung \& ttH channels can be neglected given their small cross-section.

\subsection{Heavy Neutral Lepton (HNL)}

The theoretical framework is a minimal Type-I seesaw \cite{Brdar:2019} that extends the SM by $\mathcal{N}$ right-handed neutral fermions $N_i$ with Majorana masses. The Lagrangian is
\begin{equation}
  \mathcal{L}=\mathcal{L}_{\rm SM}
  +\frac{i}{2}\,\overline{N}_i \slashed{\partial} N^i
  -\frac{m_{N_i}}{2}\,\overline{N}^i N_i^c
  -y_{i\alpha}\,\overline{N}^i \tilde{\phi}\,L^\alpha + \text{h.c.},
  \label{eq:HNL_Lagrangian}
\end{equation}

with $\alpha\in\{e,\mu,\tau\}$, $L^\alpha$ is the SM lepton doublet, $\phi$ the Higgs doublet and $\tilde{\phi}=i\sigma_2\phi^\ast$.  
We consider one HNL mixing ($i=1$) with a single flavour (one $\alpha$). After EWSB with $\langle\phi\rangle=(0,v/\sqrt{2})^T$, a Dirac mass $(m_D)_{1\alpha}=y_{1\alpha}v/\sqrt{2}$ is generated and the $(\nu_\alpha^L,N_1^c)$ mass matrix reads
\begin{equation}
  M=\begin{pmatrix}
    0 & m_D \\
    m_D & m_{N_1}
  \end{pmatrix}.
\end{equation}
For $m_{N_1}\gg m_D$, the Type-I seesaw yields $m_{\nu_\alpha}\simeq m_D^2/m_{N_1}$ and a heavy state with $m_{N}\simeq m_{N_1}$. The active–sterile mixing \cite{Li_2023} is
$
  |U_{1\alpha}|^2 \equiv \sin^2\theta_{1\alpha}=\left(\frac{m_D}{m_{N_1}}\right)^2
  =\left(\frac{y_{1\alpha} v}{\sqrt{2}\,m_{N_1}}\right)^2 .
$
In the mass basis, the HNL couples to $W^\pm,Z$ and $h$ with mixing-suppressed interactions, e.g.
\begin{align}
  \mathcal{L}_{W,Z} &\supset \frac{g}{\sqrt{2}}\,U_{1\alpha}\,W_\mu\,
  \overline{\ell}_\alpha\gamma^\mu N
  + \frac{g}{2\cos\theta_W}\,U_{1\alpha}\,Z_\mu\,
  \overline{\nu}_\alpha\gamma^\mu N + \text{h.c.}, \\
  \mathcal{L}_h &\supset -\,\frac{m_N}{v}\,U_{1\alpha}\,h\,\overline{\nu}_\alpha N + \text{h.c.}
\end{align}
Here $\theta_W$ is the Weinberg angle and $g$ the $SU(2)_L$ coupling constant. For $m_N\ll m_W$, the inclusive width scales as (approximation from \cite{Bondarenko_2018})
\begin{equation}
  \Gamma_N \simeq \frac{G_F^2\,m_N^5}{96\pi^3}\,|U_{1\alpha}|^2\,C(m_N),
  \label{eq:hnlwidth}
\end{equation}
where $C(m_N)$ captures hadronic thresholds and QCD corrections.  
For the production of the HNL \cite{Bondarenko:2018ptm} at the Large Hadron Collider (LHC), we consider meson/baryon and $\tau$ decays ($D$, $B$~mesons dominate at low $m_N$, below $\sim\!5$\,GeV) and partonic channels:
\begin{align}
  q\bar q' \to W^{*\pm}\to \ell^\pm N,
  \quad q\bar q\to Z^*\to \nu N, \quad
  gg\to h^*/Z^*\to \nu N,\quad
  q\gamma\to q'\,\ell^\pm N,
\end{align}
with $q,q'\in\{u,d,s,c,b\}$. Selection requirements however can reweight the relative sensitivity toward the partonic production topologies such that they appear to dominate for ANUBIS.  
The possible decays products, for light $m_N$,  are \cite{Bondarenko:2018ptm}
\begin{itemize}\itemsep2pt
  \item 2-body: $N\to \ell^\pm h^\mp_{P,V}$ and $N\to \nu\,h^0_{P,V}$,
  \item 3-body: $N\to \ell_\alpha\,\nu_\beta \bar\nu_\beta$ and $N\to \nu\, f\bar f$,
\end{itemize}
where $h_{P,V}$ denotes SM (pseudo)scalar/vector mesons. All channels with at least two charged particle are, in principle, visible by the ANUBIS detector.

\section{ANUBIS detector and proANUBIS}
\label{sec:detector}

\subsection{ANUBIS detector concept}
ANUBIS is a proposed detector to search for LLPs. It is planned to be composed of Resistive Plate Chambers (RPC) modules that are tiles across the ceiling of the ATLAS cavern ($\sim$20–30\,m from the IP). The RPC modules arranged in two planes (triplet RPC) separated by around 1\,m and an additional singlet layer for track disambiguation. The layout follows the ceiling curvature and provides $\sim$2\,sr solid-angle coverage. Coincidence timing with ATLAS and  integration of ANUBIS into the ATLAS trigger system, and vice versa, would enhance background rejection from cosmics or the hadronic interactions of SM particles in the muon spectrometers and enables event-by-event correlation, using bunch crossing and Global time information from ATLAS.

\subsection{\textit{proANUBIS} demonstrator}

A reduced-scale prototype, \textit{proANUBIS}, has been operating in the ATLAS cavern since 2024 and has accumulated $\sim$104\,fb$^{-1}$ of coincident $pp$ collision data. Its goals are to serve as a proof of concept for ANUBIS, evaluate the potential detector performance, develop analysis pipelines (track/vertex reconstruction, time of flight), and an in-situ study of relevant backgrounds for ANUBIS. A triplet+singlet+doublet of RPC modules are installed and aligned towards the IP. Synchronisation with ATLAS is achieved thanks to bunch-crossing and global time matching.

\begin{figure}[!h]
  \centering
  \includegraphics[width=1\linewidth]{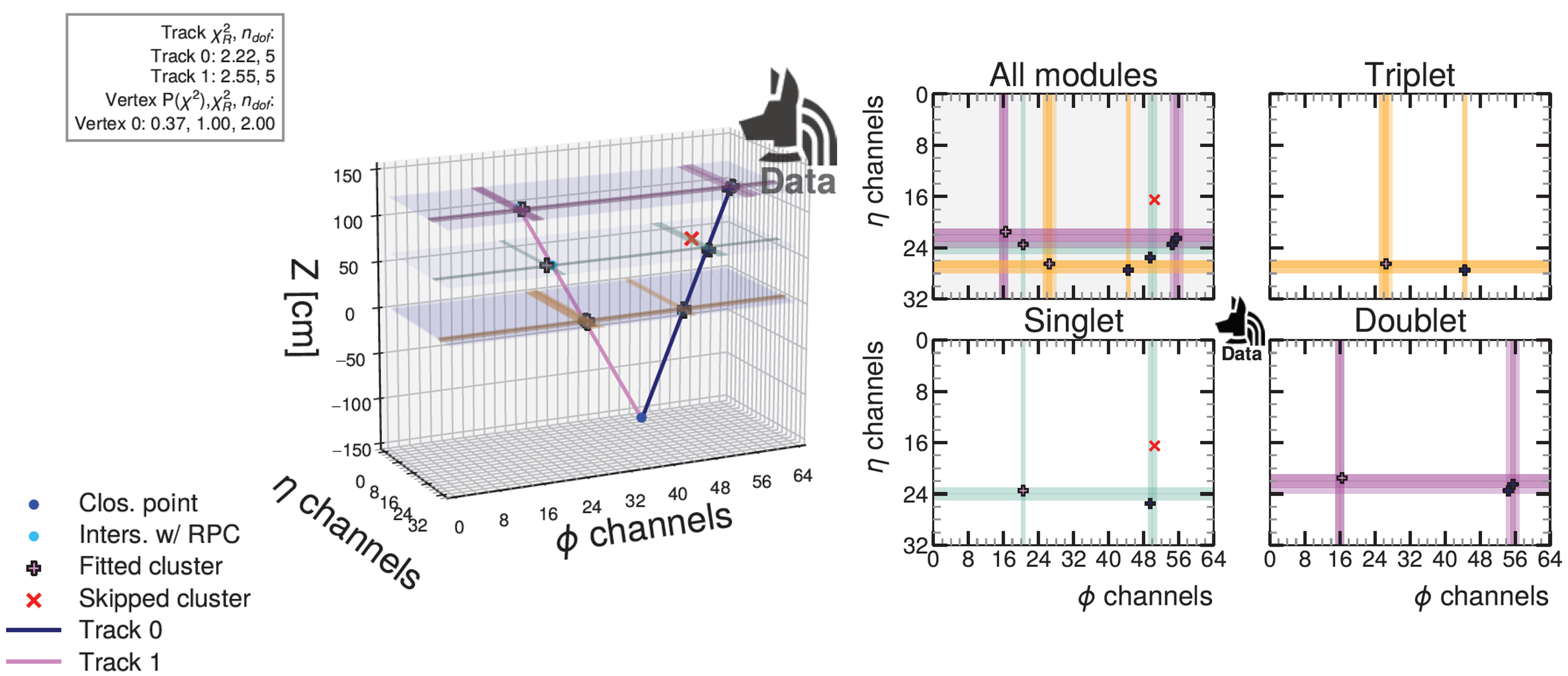}
  \caption{Example of tracks and vertex reconstruction using data from proANUBIS.}
  \label{fig:proanubis}
\end{figure}

An example of the track and vertex reconstruction is given in Fig.\ref{fig:proanubis}. The doublet and triplet are used to seed the track reconstruction, and the singlet allows for the validation of the reconstructed tracks and the disambiguation of multiple tracks in one event.

\subsection{Ongoing data analyses}

\begin{figure}[!h]
  \centering
  \includegraphics[width=1\linewidth]{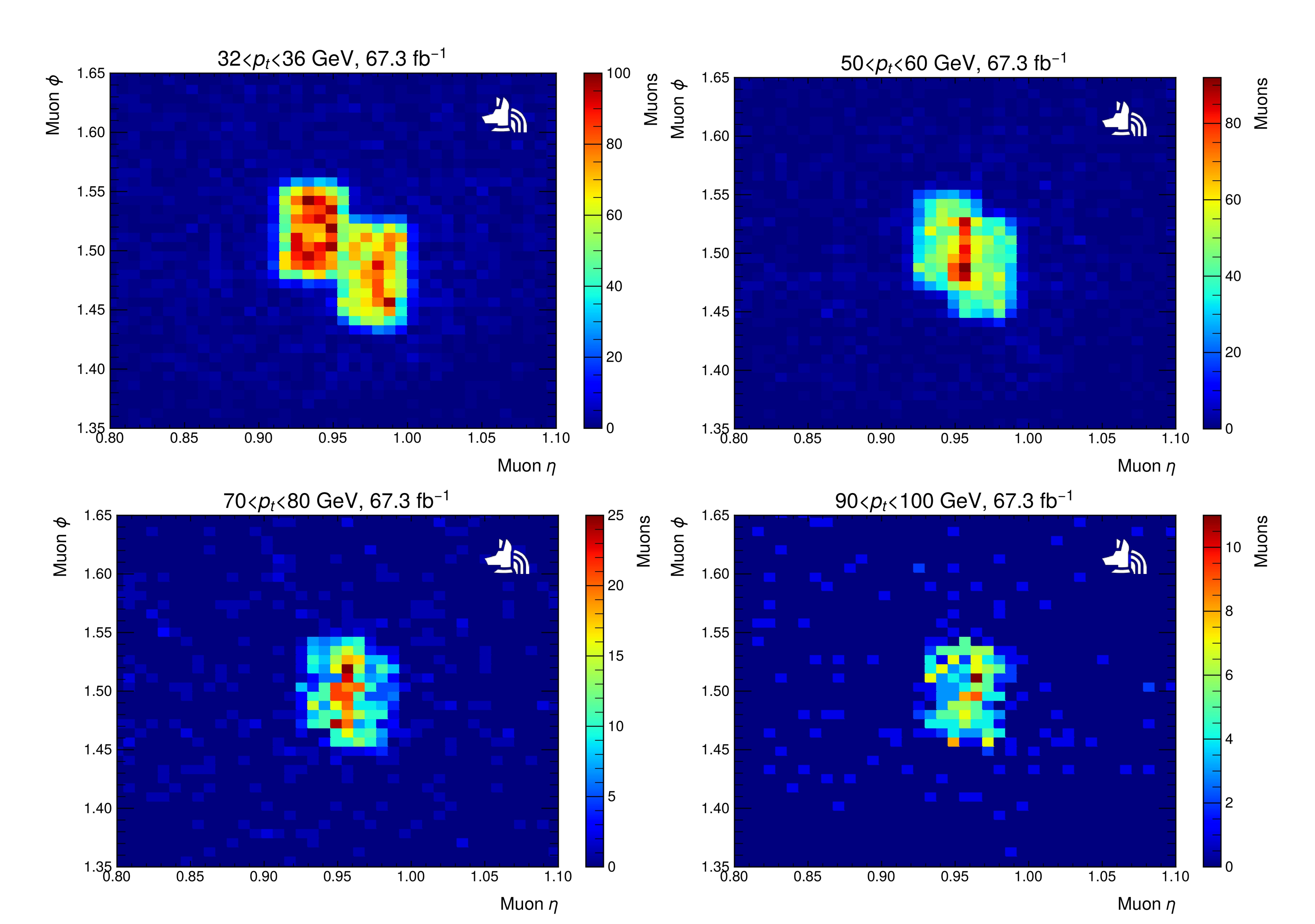}
  \caption{Heatmap of muons in synchronized events, detected by ATLAS and ANUBIS.}
  \label{fig:muons}
\end{figure}

Multiple analyses are being conducted using the proANUBIS prototype. Background estimation for the sensitivity studies, but also testing of the synchronisation with ATLAS. For example, Fig.\ref{fig:muons} shows the $\eta$ and $\phi$ coordinates of muons measured by ATLAS, that were coincident with tracks reconstructed by proANUBIS in bins of the pt of the muons in ATLAS. This effectively shows the position of proANUBIS, where the two overlapping rectangles arise due to the bending of muons and anti-muons in the MS B-field which has not been accounted for in the ($\eta$,$\phi$) projection here. At higher $pT$ the bending due to the B-field is more minimal and so the two charges converge towards the real location of proANUBIS.

\section{Framework, analysis strategy and datasets}
\label{sec:SETANUBIS}

Aiming to accelerate our future studies, we also developed \textsc{SET-ANUBIS}, a modular hexagonal framework that ingests UFO models \cite{Degrande:2011ua} and is able to compute decay widths and branching ratios (using analytic equations, \textsc{MadGraph} \cite{Alwall:2014hca} or \textsc{MARTY} \cite{Uhlrich:2021}); generate events with \textsc{Pythia8} \cite{Bierlich:2022pfr} or \textsc{MadGraph}; and then apply selection cuts based on the detector's geometry.

\begin{figure}[!h]
    \centering
    \includegraphics[width=1\linewidth]{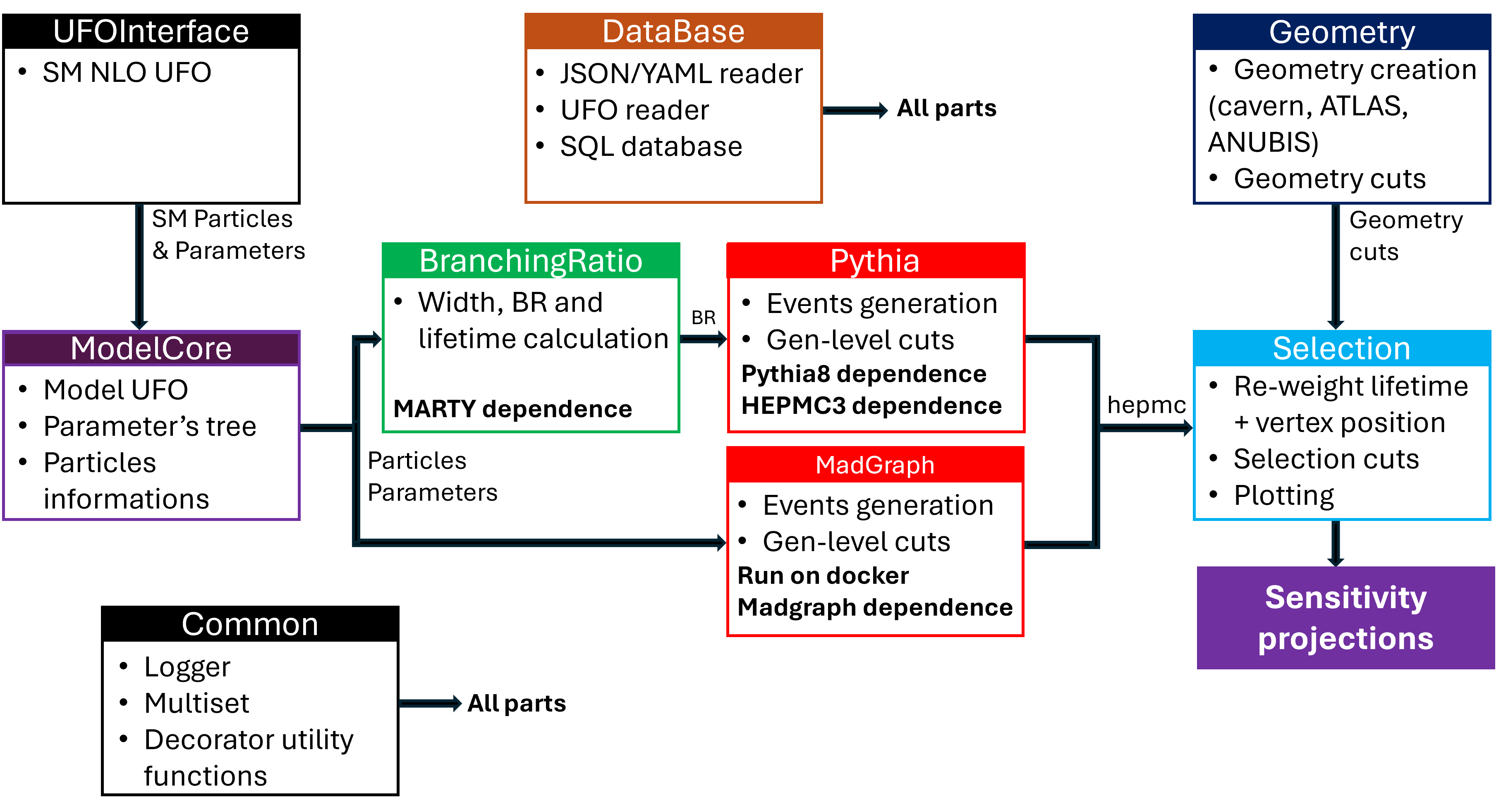}
    \caption{General architecture of SET-ANUBIS.}
    \label{fig:neosetanubisarchi}
\end{figure}

The architecture is presented in Fig.\ref{fig:neosetanubisarchi}. The main objective of this frameworks is to facilitate future studies, while being able to test different geometry and see the consequences on the detector's sensitivity.

The frameworks can use \textsc{MadGraph} or \textsc{Pythia8} for the events generation, depending on the models and the production modes. For baryonic production, MadGraph is preferred, however, for more specific production modes, such as mesonic production, \textsc{Pythia8} can be necessary.
\newpage

\section{Results}

\subsection{Higgs-portal scalar limits}

The events for calculating the Higgs sensitivity limits were generated using a ggF HAHM UFO \cite{Curtin:2015fna}. Pythia8 was then used to simulate LLP production from Higgs and decay to $b\bar{b}b\bar{b}$.

\begin{figure}[!h]
  \centering
  \includegraphics[height=8cm]{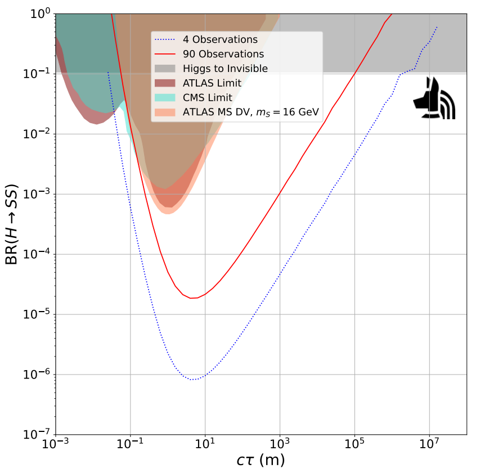}
  \caption{Projected Higgs-portal scalar exclusions for ANUBIS.}
  \label{fig:higgs}
\end{figure}

Fig.\ref{fig:higgs} shows the expected ANUBIS $2\sigma$ exclusion limits for a scalar produced via $h\to ss$ and decaying $s\to b\bar b$. The reach is driven by the large transverse decay volume. In the zero-background hypothesis, ANUBIS probes lifetime from $\mathcal{O}(\SI{1}{m})$ up to $\mathcal{O}(10^5 m)$with sensitivity to small branching ratios $ \propto 1/\mathcal{L}$. Under the conservative $B{=}90$ assumption the limit weakens as $1/\sqrt{B}$ but remains competitive thanks to geometry-driven acceptance.

\subsection{HNL limits}

\begin{figure}[!h]
  \centering
  \includegraphics[height=7cm]{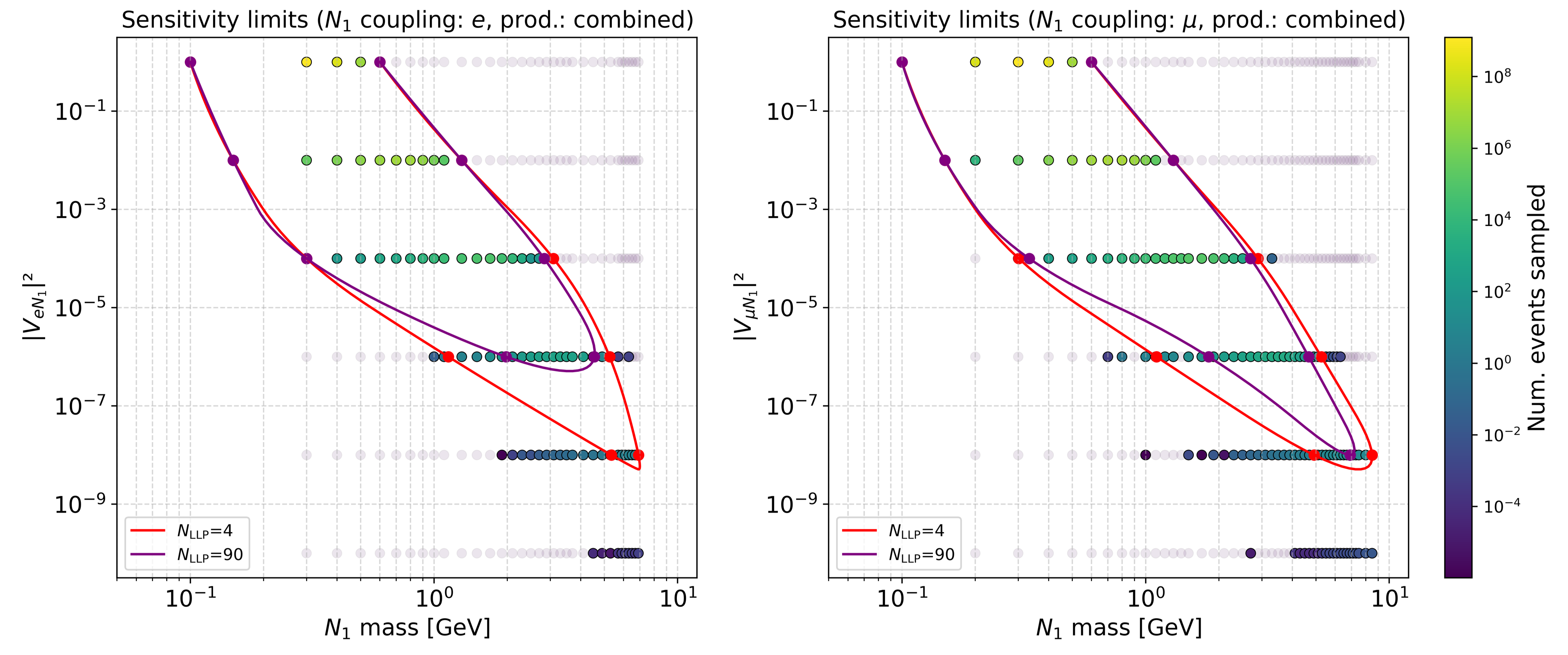}
  \caption{Projected 95\%\,CL exclusions in the $(m_N,|U_\ell|^2)$ plane for $N_e$ and $N_\mu$.}
  \label{fig:hnl}
\end{figure}

Figure~\ref{fig:hnl} presents the combined reach for $N_e$ and $N_\mu$ (one flavour at a time). Including $gg$ and $W\gamma$ channels improves acceptance for heavier, boosted HNLs. ANUBIS covers lifetimes around $10^{-1}$ to $10^3$ m scale, complementing prompt ATLAS searches and far-detector proposals. The zero- and non-zero background hypotheses bracket realistic operating scenarios.

\newpage
\newpage

\section{Conclusion}

The sensitivity studies on the HNL and Higgs-portal models and the data from the prototype show that ANUBIS offers unique sensitivity to neutral LLPs decaying in the ATLAS cavern. proANUBIS has already demonstrated its potential to take relevant data while synchronising with ATLAS and consolidating the sensitivity estimation, thanks to real background estimation. The next steps include refined background studies with ATLAS side information and extending the sensitivity estimation using \textsc{SET-ANUBIS} on additional dark-sector simplified models. Data analysis is still ongoing and will provide new results on the background estimation soon.


\begin{thebibliography}{99}

\bibitem{Hall:2010jx}
L.~J.~Hall, J.~March-Russell and S.~M.~West,
\emph{A Unified Theory of Matter Genesis: Asymmetric Freeze-In},
\href{https://arxiv.org/abs/1010.0245}{arXiv:1010.0245} [hep-ph] (2010).

\bibitem{Bondarenko:2018ptm}
K.~Bondarenko, A.~Boyarsky, D.~Gorbunov and O.~Ruchayskiy,
\emph{Phenomenology of GeV-scale Heavy Neutral Leptons},
\href{https://doi.org/10.1007/JHEP11(2018)032}{\emph{JHEP} \textbf{11} (2018) 032}
[\href{https://arxiv.org/abs/1805.08567}{\tt arXiv:1805.08567} [hep-ph]].



\bibitem{ATLAS:2013dyx}
G.~Aad \textit{et al.} [ATLAS Collaboration],
\emph{Standalone vertex finding in the ATLAS muon spectrometer},
\href{https://doi.org/10.1088/1748-0221/9/02/P02001}{\emph{JINST} \textbf{9} (2014) P02001}
[\href{https://arxiv.org/abs/1311.7070}{\tt arXiv:1311.7070} [physics.ins-det]].

\bibitem{Bauer:2019vqk}
M.~Bauer, O.~Brandt, L.~Lee and C.~Ohm,
\emph{ANUBIS: Proposal to search for long-lived neutral particles in CERN service shafts},
\href{https://arxiv.org/abs/1909.13022}{arXiv:1909.13022} [physics.ins-det] (2019).

\bibitem{Alpigiani:2018}
C.~Alpigiani \textit{et al.},
\emph{A Letter of Intent for MATHUSLA: a dedicated displaced vertex detector above ATLAS or CMS},
\href{https://arxiv.org/abs/1811.00927}{arXiv:1811.00927} [physics.ins-det] (2018).

\bibitem{Aielli_2020}
G.~Aielli \textit{et al.},
\emph{Expression of interest for the CODEX-b detector},
\href{https://doi.org/10.1140/epjc/s10052-020-08711-3}{\emph{Eur.\ Phys.\ J.\ C} \textbf{80} (2020) 1177}
[\href{https://arxiv.org/abs/1911.00481}{\tt arXiv:1911.00481} [hep-ex]].

\bibitem{Beacham_2019}
J.~Beacham \textit{et al.},
\emph{Physics beyond colliders at CERN: beyond the Standard Model working group report},
\href{https://doi.org/10.1088/1361-6471/ab4cd2}{\emph{J.\ Phys.\ G} \textbf{47} (2019) 010501}.

\bibitem{Curtin_2014}
D.~Curtin \textit{et al.},
\emph{Exotic decays of the 125 GeV Higgs boson},
\href{https://doi.org/10.1103/PhysRevD.90.075004}{\emph{Phys.\ Rev.\ D} \textbf{90} (2014) 075004}.

\bibitem{Brdar:2019}
V.~Brdar, A.~J.~Helmboldt, S.~Iwamoto and K.~Schmitz,
\emph{Type I seesaw mechanism as the common origin of neutrino mass, baryon asymmetry, and the electroweak scale},
\href{https://doi.org/10.1103/PhysRevD.100.075029}{\emph{Phys.\ Rev.\ D} \textbf{100} (2019) 075029}.

\bibitem{Li_2023}
P.~Li, Z.~Liu and K.-F.~Lyu,
\emph{Heavy neutral leptons at muon colliders},
\href{https://doi.org/10.1007/JHEP03(2023)231}{\emph{JHEP} \textbf{03} (2023) 231}.

\bibitem{Bondarenko_2018}
K.~Bondarenko, A.~Boyarsky, D.~Gorbunov and O.~Ruchayskiy,
\emph{Phenomenology of GeV-scale heavy neutral leptons},
\href{https://doi.org/10.1007/JHEP11(2018)032}{\emph{JHEP} \textbf{11} (2018) 032}
[\href{https://arxiv.org/abs/1805.08567}{\tt arXiv:1805.08567} [hep-ph]].

\bibitem{Degrande:2011ua}
C.~Degrande, C.~Duhr, B.~Fuks, D.~Grellscheid, O.~Mattelaer and T.~Reiter,
\emph{UFO -- The Universal FeynRules Output},
\href{https://doi.org/10.1016/j.cpc.2012.01.022}{\emph{Comput.\ Phys.\ Commun.} \textbf{183} (2012) 1201--1214}
[\href{https://arxiv.org/abs/1108.2040}{\tt arXiv:1108.2040} [hep-ph]].


\bibitem{Alwall:2014hca}
J.~Alwall, R.~Frederix, S.~Frixione, V.~Hirschi, F.~Maltoni, O.~Mattelaer, H.-S.~Shao, T.~Stelzer, P.~Torrielli and M.~Zaro,
\emph{The automated computation of tree-level and next-to-leading order differential cross sections, and their matching to parton shower simulations},
\href{https://doi.org/10.1007/JHEP07(2014)079}{\emph{JHEP} \textbf{07} (2014) 079}
[\href{https://arxiv.org/abs/1405.0301}{\tt arXiv:1405.0301} [hep-ph]].

\bibitem{Uhlrich:2021}
G.~Uhlrich, F.~Mahmoudi and A.~Arbey,
\emph{MARTY -- Modern ARtificial Theoretical phYsicist: A C++ framework automating theoretical calculations beyond the Standard Model},
\href{https://doi.org/10.1016/j.cpc.2021.107928}{\emph{Comput.\ Phys.\ Commun.} \textbf{264} (2021) 107928}
[\href{https://arxiv.org/abs/2011.02478}{\tt arXiv:2011.02478}].

\bibitem{Bierlich:2022pfr}
C.~Bierlich \textit{et al.},
\emph{A comprehensive guide to the physics and usage of PYTHIA 8.3},
\href{https://doi.org/10.21468/SciPostPhysCodeb.8}{\emph{SciPost Phys.\ Codeb.} \textbf{2022} (2022) 8}
[\href{https://arxiv.org/abs/2203.11601}{\tt arXiv:2203.11601} [hep-ph]].

\bibitem{Curtin:2015fna}
D.~Curtin, R.~Essig, S.~Gori and J.~Shelton,
\emph{Illuminating dark photons with high-energy colliders},
\href{https://doi.org/10.1007/JHEP02(2015)157}{\emph{JHEP} \textbf{02} (2015) 157}
[\href{https://arxiv.org/abs/1412.0018}{\tt arXiv:1412.0018} [hep-ph]].
\end{thebibliography}
\end{document}